# Nanoscale structure formation in nickel-aluminum alloys synthesized far from equilibrium

Zhehao Chen[1,*], Aslak J. J. Fellman[1], Katarzyna Mulewska[2], Kenichiro Mizohata[1], Davide Gambino[3,1], Yanling Ge[4], Eryang Lu[1], Flyura Djurabekova[1], Andreas Delimitis[5], Lukasz Kurpaska[2], Filip Tuomisto[1], and Kostas Sarakinos[1,6]

[1]*Department of Physics, University of Helsinki, P.O. Box 43, FI-00014 Helsinki, Finland*

[2]*NOMATEN Centre of Excellence, National Centre for Nuclear Research, st. A. Soltana 7, 05-400 Otwock, Poland*

[3]*Department of Physics, Chemistry and Biology (IFM), Linköping University, SE 58183 Linköping, Sweden*

[4]*VTT Technical Research Centre of Finland, P.O.Box 1000, 02044 VTT, Kemistintie 3, 02150 Espoo, Finland*

[5]Department of Physics, Aristotle University of Thessaloniki, GR-54124 Thessaloniki, Greece

[6]*KTH Royal Institute of Technology, Department of Physics, Roslagstullsbacken 21, 11421 Stockholm, Sweden*

*Corresponding author (e-mail: zhehao.chen@helsinki.fi)

**Abstract**

The present study reports on the structure formation in thin epitaxial nickel-aluminum films ($Ni_{1-x}Al_x$; Al atomic fraction $x$ up to $x = 0.24$) grown on MgO$(001)$ substrates by magnetron sputtering. Experimental and computational data demonstrate that for $x < 0.11$, the films exhibit the face-centered cubic random solid-solution $Ni_{1-x}Al_x$ structure ($\gamma$ phase). Whereas in the range $x = 0.11 - 0.24$ the $\gamma$ phase coexists with the ordered $L1_2$ structure ($\gamma'$ phase). The two phases are homogenously intermixed forming a coherent and strained *nano-solution*, which exhibits a single lattice parameter that expands as the Al content increases. Isothermal annealing of films containing $x = 0.14$ of Al, coupled with structural and nano-mechanical characterization, reveal that the nano-solution retains its overall integrity for temperatures up to $673\ K$, while the film hardness increases from $5.5\ GPa$ (as deposited films) to $6\ GPa$. Further increase of the annealing temperature to $873\ K$ and $1073\ K$ causes the nano-solution to dissolve into distinct $\gamma$ and $\gamma'$ phase domains and the hardness to decrease down to values of $4\ GPa$. These findings confirm the metastable nature of the as-deposited thin $Ni_{1-x}Al_x$ alloy films and underpin the effectiveness of high supersaturation/undercooling for creating non-equilibrium phases and self-organized nanostructures upon synthesis of multicomponent materials.

## 1. Introduction

Nickel-aluminum ($Ni_{1-x}Al_x$) alloys—most notably alloys with Al atomic fraction $x$ in the range $0 – 0.25$ —are renowned for their excellent mechanical properties and high-temperature stability, rendering them important for aerospace, automotive, and electronic applications [1-4]. The equilibrium Ni-Al phase diagram [5] predicts the formation of a random face-centered cubic (FCC) $Ni_{1-x}Al_x$ solid solution (denoted as $\gamma$ phase) for $x \leq 0.1$, while an ordered phase with the $L1_2$ crystal structure (denoted as $\gamma'$ $Ni_3Al$ phase) forms for $x \cong 0.25$. In the Al atomic fraction range $0.1 < x < 0.25$ discrete $\gamma$ and $\gamma'$ phase domains coexist, forming coherent interfaces due to their small lattice mismatch ($< 0.5\%$) [6, 7]. This microstructure further contributes to the alloy strength, creep, and corrosion resistance and has been one of the focal points of research work [8-11] pertaining to the Ni-Al system.

Alloying—a key route for developing materials with tailor-made properties within physical metallurgy—has been extensively explored through equilibrium synthesis. Concurrently, contemporary materials science and technology has been increasingly relying on non-equilibrium synthesis routes in which high cooling rates (quenching) [12], large external pressures [13], and atom-by-atom growth of thin films from the vapor phase [14] impose limited atomic-assembly kinetics that enable formation of metastable phases and configurations beyond the predictions of thermodynamics. For the case of $Ni_{1-x}Al_x$ alloys, rapid solidification has been a commonly used non-equilibrium synthesis approach [15-23]. It has been shown that variation of the quenching rates can significantly affect the microstructure and phase formation, e.g., by reducing grain size, increasing dislocation density, and generating fine antiphase domain (APD) structures [15, 18, 20, 24] that enhance ductility [15, 25], the lack of which is one of the main challenges faced by contemporary $Ni_{1-x}Al_x$ alloys [2, 3].

Physical vapor deposition of thin films is an effective way to synthesize metastable solid solutions and self-organized nanostructures. Examples include metal nitride, carbide, and boride multinary systems, [2, 26-30] which exhibit (among other features) unique mechanical properties and high-temperature stability [31, 32] Magnetron sputtering [14, 33, 34], a physical vapor deposition technique that has rapidly developed since the 1980s, is now well-established for industrial production of films and coatings. It is also a very far-from-equilibrium process offering cooling rates of $\cong$ $10^{14}$ $\mathrm{Ks}^{-1}$ [35] upon vapor condensation, which are several orders of magnitude larger than those achieved by quenching. The effect of such extreme cooling rates on the structure formation of $Ni_{1-x}Al_x$ alloy films, its potential for leading to metastable phases and configurations, and its implications for thermal stability and mechanical properties remain largely unexplored.

In the present work, we set out to study structure formation, thermal stability, and mechanical strength of sputtered-deposited thin $Ni_{1-x}Al_x$ films with Al atomic fractions $\mathrm{x}$ in the range $0$ (pure Ni) to $0.24$. We elect to synthesize epitaxial layers on MgO$(001)$ substrates, such that potential effects of grain boundaries and other extended defects on phase formation and stability are mitigated. Using structure and chemistry probes, along with theoretical calculations, we show that for Al fractions $\mathrm{x} < 0.11$ a random FCC $Ni_{1-x}Al_x$ solid solution ($\gamma$ phase) forms. Whereas in the range $0.11 \leq \mathrm{x} \leq 0.24$, a *nano-solution* is formed between strained $\gamma$- and $\gamma'$-phase domains which are homogeneously intermixed at atomic scale. This nano-solution exhibits a single and common to both phase domains lattice parameter that progressively expands with increasing Al content. Mechanical properties evaluation shows that the film hardness increases from $3.1$ GPa to $5.$ GPa with increasing Al fraction from $\mathrm{x} = 0.07$ to $\mathrm{x} = 0.24$. Annealing experiments on samples with $\mathrm{x} = 0.14$ show that the nano-solution structure remains largely intact, while the hardness increases from $5.5$ GPa for the as deposited films to $6$ GPa for the annealed films. Further annealing up to $1073$ K causes decomposition into discrete $\gamma$- and $\gamma'$-phase domains and softening down to a hardness value of $4\ \mathrm{GPa}$. We argue that the initial annealing-induced hardening is caused by enhancement of compositional modulation and lattice strain, while the softening is associated with domain coarsening when the structure separates into distinct $\gamma$ and $\gamma'$ phases. The overall results of our study are a further confirmation that non-equilibrium thin-film deposition is an effective route for synthesizing metastable nanostructures that extend beyond systems comprising single-phase compounds into the realm of complex intermetallics.

## 2. Research methodology

### 2.1. Thin-film synthesis and characterization

$Ni_{1-x}Al_x$ layers with various Al atomic fractions $\mathrm{x}$ are deposited epitaxially by magnetron co-sputtering on $10 \times 10 \times 1\ \mathrm{mm}^3$ MgO$(001)$ substrates in an ultra-high vacuum deposition system (base pressure $\cong 10^{-6}\ \mathrm{Pa}$), equipped with a load-lock chamber and three magnetron sources. The substrates are cleaned with trichloroethylene, acetone, and ethanol for 10 minutes each and are dried with $N_2$. Immediately after cleaning, the substrates are mounted on a Mo holder, introduced to the deposition chamber, annealed to $1073\ \mathrm{K}$ for $60$ minutes, and subsequently cooled to the deposition temperature of $473\ \mathrm{K}$. The substrate is heated using radiation from a resistive heater placed less than $5\ \mathrm{mm}$ behind the substrate holder. Since the MgO substrate is transparent, a tungsten plate is positioned behind the substrate to absorb the radiation generated by the resistive heater and transfer it via direct contact with the substrate. The substrate temperature is monitored by a thermocouple placed between the substrate and the heater.

Ar ($99.99\%$ purity) is used as the sputtering gas, and the discharge pressure is maintained at $0.66\ \mathrm{Pa}$ during deposition. Ni and Al targets (purity $99.995\%$; diameter $50.8\ \mathrm{mm}$) are used for the deposition process. Radio frequency (RF) power ($150\ \mathrm{W}$) is applied on the Ni target, while the Al target is operated in direct current (DC) mode with the power ranging from $0$ to $51\ \ \mathrm{W}$ is to deposit layers with different Al contents. The substrate holder is rotated at a constant speed of $3\ \mathrm{rpm}$ to obtain uniform deposition over the substrate area. Under the synthesis conditions outlined above, and with a target-to-sample distance of $190\ \mathrm{mm}$, the deposition rate is in the range $0.06 - 0.08\ \mathrm{nm\ s^{-1}}$, as measured by a quartz crystal microbalance (QCM) and confirmed by transmission electron microscopy (TEM). The deposition time is $3000\ \mathrm{s}$ to achieve film thicknesses in the range $180\ -\ 240\ \mathrm{nm}$.

The global elemental composition of the films is determined by time-of-flight elastic recoil detection analysis (ToF-ERDA) in the EGP-10-II 5 MV tandem accelerator of the Helsinki Accelerator Laboratory [36]. Measurements are performed using a $50\ \mathrm{MeV}$ $^{127}$I ion beam, with the detector positioned at an angle of $40°$ with respect to the direction of the incident beam, while the beam incident direction forms an angle of $20°$ with respect to the sample surface. Compositions are calculated using stopping forces obtained from the SRIM freeware [37], the measurement geometry, and Rutherford elastic recoil cross sections for the detected elements.

High-resolution x-ray diffractometry (XRD) measurements are conducted with a Rigaku SmartLab X-ray diffractometer, employing Cu $K_{\alpha,\beta}$ radiation and a Ge (220) monochromator for studying the film phase composition and crystal structure. Measurements performed in $\theta - 2\theta$ geometry ($2\theta$ range 10-90°) show that all layers grow along the $< 100 >$ direction exhibiting maxima that correspond to the $Ni_3Al(001)$, $Ni_3Al(002)$ ,and Ni $(002)$ reflections (see Section 3 for details). Based on those reflections, the out-of-plane lattice parameters of the various phases are computed. XRD measurements are also performed in rocking curve ($\omega - \mathrm{scan}$) geometry around the $(002)$ reflection of the layers, while pole figures are collected with respect to the positions of the $(111)$ and $(002)$ $Ni_{1-x}Al_x$ layer reflections.

A JIB-4700F focused ion beam (FIB) system, capable of performing scanning electron microscopy (SEM), energy dispersive x-ray spectroscopy (EDS), and electron backscatter diffraction (EBSD), is utilized to examine the film surface morphology, determine elemental distribution, and prepare cross-sectional lamellae for transmission electron microscopy (TEM) analyses [38, 39]. The TEM lift-out procedure is performed on the surface region with the area of interest covered by a platinum (Pt) protective layer to prevent further ion damage before lamella lift-out. The lamella is subsequently thinned to approximately $100\ \mathrm{nm}$ by ion milling and polishing and is attached to a molybdenum grid for high-temperature annealing preparation. The coarse milling process employs a $30\ \mathrm{keV}$ galium (Ga) ion beam with a current of $500\ \mathrm{pA}$, followed by a fine polishing stage using a $5\ \mathrm{keV}$ Ga ion beam with a current of $30\ \mathrm{pA}$ to remove the amorphous layer and additional damage introduced during thinning.

The general microstructure, crystalline quality, and phase composition of the films, as well as the elemental distribution therein, are investigated using high-resolution TEM (HRTEM; spatial resolution 0.1nm at $200\ \mathrm{kV}$), TEM large angle energy dispersive spectrometer (EDS), selected area diffraction (SAD), and complementary bright filed (BF) and dark field (DF) imaging in a JEOL JEM-2800 instrument. To facilitate accurate interpretation of HRTEM data with respect to the atomic positions and crystal symmetry, we perform image simulations using the AbTEM algorithm [40].The simulation sample is constructed by creating a two-phase sample comprising both $L1_2$ $Ni_3Al$ and FCC Ni structures with dimensions $114 \times 57 \times 890\ \mathrm{nm}^3$. The simulation parameters are chosen to reflect the experimental setup and listed in the Supporting Information (Section S1)[41].

The reduced modulus $Y$ and hardness $H$ of the $Ni_{1-x}Al_x$ layers are determined by nanoindentation (NI). Measurements are performed using a NanoTest Vantage system (Micro Materials Ltd.) equipped with a Berkovich-shaped diamond indenter tip. $Y$ and $H$ values are extracted from nanoindentation load-displacement curves using the formalism developed by Oliver and Pharr [42]. To assess the effect of substrate on the measured film mechanical properties, experiments are conducted in single-force mode with the depth control method, whereby penetration depths are ranging from $40$ to $100\ \mathrm{nm}$. For each indentation depth, $20$ measurements are performed, maintaining a $20\ \mu\mathrm{m}$ distance between indents to avoid interference between neighboring indentation sites.

Selected $Ni_{1-x}Al_x$ samples are subjected to in-vacuum isothermal annealing (using the deposition chamber) with the purpose of studying the thermal stability of the film structure and crystalline phases. Annealing is performed at temperatures of $673\ \mathrm{K}$, $873\ \mathrm{K}$, and $1073\ \mathrm{K}$ for one hour with the base pressure being in the range of $10^{-4}$ to $10^{-5}$ Pa, followed by XRD measurements. The samples also undergo NI analysis to correlate annealing-induced changes in the film microstructure with their mechanical properties.

### 2.2. Theoretical calculations

To further understand the evolution of the lattice parameter of the $Ni_{1-x}Al_x$ layers as a function of the Al content $x$, we perform $0$ K density functional theory (DFT) and molecular statics (MS) calculations for random substitutional solid solutions ($\gamma$ phase) and ordered $\gamma'$-phases (both off-stoichiometric and stoichiometric) in the $x$ range $0$ to $0.25$. Random solid solution structures are created by using a pure Ni FCC cell as a starting point and randomly replacing Ni by Al atoms. Ordered systems are created starting from the $L1_2$ $Ni_3Al$ structure and replacing Al by Ni atoms. This procedure allows us to control the Al concentration, while maintaining local atomic ordering in the $L1_2$ structure. The lattice parameters are estimated by calculating the energy of the system at different volumes and computing the equilibrium lattice constant via the Murnaghan equation of state [43].

DFT calculations are carried out for cells containing $108$ atoms arranged either in FCC random $Ni_{1-x}Al_x$ solid-solutions or in the $L1_2$ $Ni_3Al$ structure. Due to system size limitations, Al-deficient $Ni_3Al$-type cells are not considered for DFT calculations. All calculations are performed within the Vienna Ab Initio Simulation Package (VASP)[44-46]. We use the PBE GGA exchange-correlation function [47] and projector augmented wave (PAW) pseudopotentials (Al, Ni_pv). The plane-wave expansion energy cutoff is $650$ eV for all elements. K-points are defined using $\Gamma$-centered Monkhorst–Pack grids with maximum k-spacing of $0.15$ $A^{-1}$. Moreover, we apply a first order Methfessel-Paxton smearing of $0.1$ eV [48].

MS calculations are performed within the Large-Scale Atomic/Molecular Massively Parallel Simulator (LAMMPS) [49]. We use cells that consist of $4000$ atoms and we compute interatomic forces using a recently developed machine-learned potential for AlCrCuFeNi [50]. The interatomic potential is based on the Gaussian approximation method [51] and it is a low dimensional tabulated version (tabGAP) thereof [52, 53]. The model uses an external repulsive potential, along with a two-body descriptor, a three-body descriptor, and an embedded-atom-method-like descriptor. The parameters of the formalism are determined by training the model on an extensive dataset of DFT calculated structures that includes the constituent binary alloys (such as the $Ni_{1-x}Al_x$ in the present work). We note here that the potential is based on elemental Ni and Al data from recent single element tabGAP models [54]. Initial tests showed that the original formulation of the interatomic potential underestimates the lattice constants of the $Ni_{1-x}Al_x$ alloys with reference to DFT calculations. This is to be expected since the potential is not explicitly designed for the binary Ni-Al system. Hence, the model has been improved to match DFT lattice constant predictions by modifying its hyperparameters. The list of the hyperparameters used herein are presented in the Supporting Information (Section S2) [41].

## 3. Results and discussion.

### 3.1 Crystal structure and phase formation

Figure 1 presents the x-ray $\theta - 2\theta$ diffraction patterns (intensity in logarithmic scale) of Ni and $Ni_{1-x}Al_x$ films. The Al content is varied by changing the power applied to the Al sputtering cathode, as shown in Supporting Information (Section S3 and Fig.S1) [41]. Panel (a) shows patterns in the $2\theta$ range $10 - 90°$, along with the angular position of key reflections for the bulk unstrained NaCl-type MgO (ICDD File: #00-045-0946), FCC Ni (ICDD File: #00-004-0850), and $L1_2$ $Ni_3Al$ [6], which are marked with vertical dashed lines. The MgO$(002)$ peak emanating from the substrate is observed in all patterns. Moreover, two additional peaks are visible: (i) for all samples a peak at an approximate angular position of $52°$ in close proximity to the Ni $(002)$ and $Ni_3Al$ $(002)$ reflections; and (ii) for $Ni_{1-x}Al_x$ films with $x \geq 0.11$ a peak at an approximate angular position of $25°$ closely matching the $Ni_3Al$ $(001)$ reflection. Since no other peaks are observed in Fig.1 (a), we conclude that our layers exhibit an out-of-plane texture along the $< 100 >$ crystallographic orientation. This finding, along with pole figure and rocking curve measurements (representative data for Ni are shown in Section S4/Fig.S2 in the Supporting Information [41]) establish that our Ni and $Ni_{1-x}A_x$ layers grow epitaxially on MgO$(001)$ substrates with orientation relationships $(001)_{Ni} \parallel (001)_{MgO}$; $[100]_{Ni} \parallel [100]_{MgO}$ and $(001)_{Ni_{1-x}Al_x} \parallel (001)_{MgO}$; $[100]_{Ni_{1-x}Al_x} \parallel [100]_{MgO}$. The notion of this epitaxial relationship is also supported by the SAD patterns in Fig.2 (see Section 3.2 for details). Epitaxial growth of Ni on MgO$(001)$ has been previously observed by Nakai *et al.* [55] and Milosevic *et al.* [56]. This is despite the large lattice mismatch $(17\%)$ between the film and the substrate and has been explained by the formation of a periodic array of misfit dislocations at the film-substrate interface. It is likely that a similar mechanism also facilitates epitaxial growth of $Ni_{1-x}Al_x$ films, however detailed study of the interface structure and its effect on film growth is beyond the scope of the present work.

To further understand the effect of Al atomic fraction on the $Ni_{1-x}Al_x$ film crystal structure and phase composition, magnified areas in the $2\theta$ angular ranges of $23.5°–26.5°$ and $49°–54°$ are presented in Fig.1 panels (b) and (c), respectively. Figure 1(b) confirms that the $Ni_3Al(001)$ peak only appears for $x \geq 0.11$. Furthermore, with increasing Al content the peak marginally shifts from $2\theta = 25.06°$ at $x = 0.11$ to $2\theta = 25.0°$ at $x = 0.24$. Concurrently, the Al content increase causes the peak to become sharper, with the full width at half maximum (FWHM) decreasing from $0.62°$ to $0.26°$. Turning to Fig.1(c), we observe that the peak gradually shifts from the Ni$(002)$ position at $2\theta = 51.91°$ for $x = 0$ to nearly the $Ni_3Al(001)$ position $(2\theta = 51.24°)$ for the $Ni_{0.76}Al_{0.24}$ sample, while remaining symmetric. From the results presented in Fig.1, we conclude that for $x < 0.11$ the films consist of the FCC $Ni_{1-x}Al_x$ random solid solution $\gamma$ phase, while the $\gamma$ and $\gamma'$ $(L1_2)$ phases seemingly co-exist in the films for $x \geq 0.11$.

XRD data are complented by TEM analyses. Figure 2 shows cross-sectional TEM images and SAD patterns of the $Ni_{1-x}Al_x$ thin films for $x = 0, 0.05, 0.14, \text{and } 0.24$. The BF images in Figs.2(a)-(d) illustrate the overall structure of the films and the MgO substrate (film thicknesses ranging from $\cong 100$ to $\cong 200$ nm), from which we establish that the film structure is uniform, and the film-substrate interface is sharp/well defined, with no signs of extended defected regions. Figures 2 (a1) and (a2) show the SAD patterns of the Ni films and the Ni/MgO interface along the $[100]$ zone axis. The patterns exhibit square symmetry (consistent with the FCC and NaCl-type structure of Ni and MgO). In Fig.2(a2) the corresponding reflections from both Ni and MgO are marked by dashed white and red squares, respectively. These two patterns exhibit coherent alignment, indicating high-quality epitaxial growth of the film on the MgO substrate with a perfect cube-on-cube orientation relationship. Similar features/conclusions are observed/drawn from the SAD patterns for the $Ni_{0.95}Al_{0.05}$ (panels (b1) and (b2)), $Ni_{0.86}Al_{0.14}$ (panels (c2) and (c3)), and $Ni_{0.76}Al_{0.24}$ (panels (d2) and (d3)) samples. The overall conclusions with respect to epitaxial growth established from the TEM data in Fig.2 agree with those from the XRD data in Figs.1 and S2.

Further analysis of the TEM data shows that additional superlattice spots are present in the SAD patterns for the $Ni_{0.86}Al_{0.14}$ and $Ni_{0.76}Al_{0.24}$ samples (the $(10\bar{1})$ spot is marked with dashed white circles in panels (c2) and (d2)) which indicates formation of the $L1_2$ structure (as also seen for the same Al content range in Fig.1). The marked superlattice reflections are used to acquire DF imaging with the objective of identifying areas of segregated $\gamma'$-$Ni_3Al$ phase, as such segregation is commonly observed in metallurgically synthesized alloys [57]. DF images are displayed in Figs.2(c1) and (d1) showing a uniform structure with no large segregation or phase separation. This finding correlates well with STEM-EDS data that show uniform distribution of Ni and Al in both $Ni_{0.86}Al_{0.14}$ and $Ni_{0.76}Al_{0.24}$ samples (Figs. 2(c4)-(c6) and (d4)-(d6). The overall conclusion of the data pertaining to the $Ni_{0.86}Al_{0.14}$ and $Ni_{0.76}Al_{0.24}$ samples in Fig.2 is that potentially coexisting $\gamma$ and $\gamma'$ phases are distributed uniformly in the film at the micrometer scale.

### 3.2 Nanoscale structure formation

High resolution TEM (HRTEM) is performed with the purpose of establishing a nanoscale picture of the phase composition within the $Ni_{1-x}Al_x$ films. HRTEM images from $Ni_{1-x}Al_x$ samples for $x = 0, 0.05, 0.14, \text{and } 0.24$, are presented in Figs.3(a), (b), (c), and (d), respectively, along with the corresponding fast Fourier transform (FFT) as insets. In all images, no clear grain boundary is observed, and the lattice fringes show high coherency. In the Ni (panel (a) and $Ni_{0.95}Al_{0.05}$ (panel (b) samples, the atomic layers are homogeneous, and the difference between $(001)$ and $(002)$ atomic planes is indistinguishable. The FFTs show only the spots corresponding to the FCC structure, indicating that the $\gamma'$ phase is not significantly present at this low Al content. The images corresponding to the $Ni_{0.86}Al_{0.14}$ (panel (c)) and $Ni_{0.76}Al_{0.24}$ (panel (d)) films show distinct $(001)$ and $(002)$ atomic planes with dark and bright lattice fringes and superlattice spots from the $L1_2$ structure in the FFT, suggesting the formation of the $\gamma'$ phase. For the case of the $Ni_{0.86}Al_{0.14}$ film, the $(002)$ fringes are intermittently visible, indicating that the $\gamma'$ phase is highly mixed with the $\gamma$ phase. The domain size of each phase is very small, irregular, and not well defined, i.e., the two phases mixed at the atomic scale. In contrast, the image for the $Ni_{0.76}Al_{0.24}$ film—which is very close to the stoichiometric $Ni_3Al$ composition—shows a much more distinct and uniform $\gamma'$ phase. The atomic layers are more ordered, with clear and continuous $(001)$ and $(002)$ fringes, indicating less intermixing with the $\gamma$ phase compared to the $Ni_{0.86}Al_{0.14}$ sample. This is also supported by the superlattice spots in the FFT pattern, which are brighter and sharper.

In Fig.3(c), a typical region that contains both $\gamma'$ and $\gamma$ phases is marked with red rectangle and is enlarged in Fig.3(c3) for further analysis. The image set of Fig.3(c1) through (c4) shows the supercell model (panel (c1)) used for image simulation (see Section 2 for details), the corresponding simulated HRTEM image (panel (c2)), the experimental HRTEM image (panel (c3)), and the comparison of the experimental and simulated intensity profiles obtained across the $\gamma'/\gamma$ interface from experimental and simulated images (panel (c4)). The simulated HRTEM image using mixed $\gamma'$ and $\gamma$ supercell model matches well with the experimental HRTEM image (Fig.3(c1)-(c3)), which provides evidence in favor of a coherent dispersion of the two phases (marked in yellow and white rectangular boxes in panels (c2) and (c3)). Similar structures were recently reported by Yong-Qiang *et al.* [58], in a VCoNi-W-Cu-Al-B alloy, where they described them as supranano $L1_2$ particles embedded in an FCC matrix.

The intensity profiles in Fig.3(c4) show that the $(001)$ interplanar spacing is the same for both phases. This provides further support to the notion that the crystal lattice is coherent across the various phase domains and exhibits a single lattice parameter. Moreover, a distinct transition zone from the $\gamma'$ phase to the $\gamma$ phase is observed, measuring approximately $1.40$ nm, equivalent to about eight atomic $(002)$ planes. This width is in agreement with our image simulations and very close to the thermal equilibrium value of $1.23 \pm 0.19$ nm reported by Forghani *et al.* [7]. In this transition zone, the lattice is distorted, and a coherent structure forms by contraction/expansion of the $\gamma'/\gamma$ phase lattice planes. For a unit cell of either phase to relax to its unstrained lattice parameter, the domain radius should be much larger than the transition zone of $1.4$ nm. This is not the case as seen in Fig.3(c) where both $\gamma'$- and $\gamma$-phase domains are ultra-fine and intermixed at the atomic scale.

We also observe that the HRTEM image in Fig.3(c3) exhibits a distinct structural feature marked by the white dashed rectangular box. This feature can be explained as atomic phase misalignment, where part of the $\gamma'$ phase shifts along the $\frac{1}{2}[01\bar{1}]$ direction, leading to an anti-phase boundary that is perpendicular to the beam direction $[100]$. This is a common type of planar defect in $\gamma'$-$Ni_3Al$ phase [59], whereby the original $\gamma'$ phase pattern overlaps with the shifted $\gamma'$ phase pattern.

The overall evolution of the XRD diffraction patterns in Fig.1, along with the HRTEM data in Fig.3, indicate that for Al atomic fractions $x < 0.11$ the films form a γ-phase $Ni_{1-x}Al_x$ random solid-solution, while for $x \geq 0.11$ the γ and the ordered γ'phases coexist.

Moreover, Fig.1 shows that the shape of the $(002)$ peak between the Ni$(002)$ and $Ni_3Al$ $(002)$ reflections remain symmetric without visible sign of splitting or deconvolution into multiple peaks, i.e., a single coherent lattice is formed, in agreement with the HRTEM data for the $Ni_{0.86}Al_{0.14}$ sample in Fig.3. Concurrently, the $(002)$ peak broadens for intermediate values in the $x$ range $0$ to $0.24$ relative to the peak shape at the extreme $x$ values, i.e., $\mathrm{FHWM}$ is equal to $0.28°$, $0.42°$, and $0.36°$ for Al atomic fractions of $0$, $0.17$, and $0.24$, respectively. This is indicative of lattice distortion and strain when both phases coexist.

We now return to the XRD data in Fig.1. and use the angular positions of the $(001)$ and $(002)$ reflections to calculate the out-of-plane lattice parameters of the $Ni_{1-x}Al_x$ layers. The results are plotted as a function of the Al content in Fig.4, whereby blue hollow triangles and black hollow circles correspond to data from the $(001)$ and $(002)$ XRD reflections, respectively. The lattice parameter increases in a nearly linear fashion up to an Al fraction of x=0.11 above which the increase becomes seemingly sub-linear. Lattice parameter data extracted from MS calculations for $Ni_{1-x}Al_x$ alloys with the random solid-solution γ (blue hollow squares) and the ordered γ' (red hollow squares) structures are also plotted in Fig.4, along with data from DFT calculations (blue and red hollow stars). The results from the two computational approaches compare favorably, which lends confidence to the accuracy of the machine-learned interatomic potential used for capturing the lattice parameter variations as function of Al atomic fraction. The lattice parameters for the γ and γ' structure coincide up to an Al fraction of x≅0.1, above which they deviate with the lattice parameter of the $\gamma$ structure growing in a super-linear fashion and its $\gamma'$ counterpart exhibiting a sub-linear evolution. Comparison of experimental and computational data show that for $x < 0.11$ the experimental lattice parameter follows the trend of the calculated values for the $\gamma$ and $\gamma'$ phases, while for $x \geq 0.11$ the experimental lattice parameters converge towards the computed value for the off-stoichiometric $L1_2$ γ' structure. At x=0.24, the experimental lattice parameter is very close to that of the $Ni_3Al$, indicating that the film predominantly consists of $\gamma'$ phase $Ni_3Al$.

The fact that a single lattice parameter is measured when both $\gamma$ and $\gamma'$ domains coexist suggests that either the two phases have the same equilibrium lattice parameter or are strained to adopt a common average value. According to the computed lattice parameters in Fig.4, if the Al content $x$ in both $\gamma$ and $\gamma'$ phases correspond to the global alloy composition value, their respective lattice parameters should differ. Their values begin to diverge for $x \geq 0.1$, with $\gamma$ exhibiting a larger lattice parameter than $\gamma'$, while their difference grows with increasing $x$. Hence, achieving identical equilibrium lattice parameters for both $\gamma$ and $\gamma'$ phases would require fine tuning of the Al content according to the lattice parameter vs. composition curves plotted in Fig.4, along with fine tuning of the domain size to achieve a specific global alloy composition.

Although our results do not unequivocally rule out the above-explained scenario, we propose a more likely pathway in which $\gamma$ and $\gamma'$ domains have an Al content close to the global composition of the alloy and different lattice parameters. The phase domains are then intermixed at the nanoscale forming a strained coherent lattice. In this structure the strain field emanating at the coherent interfaces between the $\gamma$ and $\gamma'$ phases does not decay within the nanometer-size domains and relaxation of the lattice to the corresponding equilibrium values of the two phases is not possible. This results in continuous local variation of the lattice constant around an average value between those of $\gamma$ and $\gamma'$ phases. The average lattice parameter expands as the global Al atomic fraction $\mathrm{x}$ increases, with the $\gamma$ phase being progressively substituted by the $\gamma'$ phase. This behavior is similar to what is observed in a substitutional solid solution, but in our case the substitution and mixing happen at the level of nanometer-size phase domains. We call this structure *nano-solution*. The formation of the nano-solution is supported by the intermixed $\gamma$ and $\gamma'$ domains within the transition zone in Fig.3(c4), the increase of the FWHM of the $(002)$ XRD reflections for $\mathrm{x} = 0.17$ in Fig.1(c), and the hardness evolution vs. composition as discussed in Section 3.3. The fact that the $\gamma$ and $\gamma'$ exhibit a nanometer-scale size indicate that structure formation may proceed via surface-directed spinodal decomposition [60-63].

### 3.3 Phase stability and mechanical properties

The nano-solution structure established by the data presented in Sections 3.1 and 3.2 is potentially metastable due to the limited diffusion of Ni and Al adatoms caused by the ultrafast cooling rates during vapor condensation in the sputtering process. To explore the viability of this hypothesis, a sample with composition $Ni_{0.86}Al_{0.14}$ is annealed at $673$, $873$, and $1073\ \mathrm{K}$. The XRD patterns, reduced modulus and hardness (the latter two determined via NI), and surface morphology (studied by SEM) of the as-deposited $(473\ \mathrm{K})$ and annealed samples are presented in Fig.5.

With reference to the as-deposited sample, the XRD patterns (Figs.5(a1)–(a3)) reveal that annealing at $673\ \mathrm{K}$ causes the $(001)$ peak (panel (a1)) to become broader $(\mathrm{FWHM} = 0.84°$ vs. $\mathrm{FWHM} = 0.65°$ for the as-deposited sample) and shift slightly to the left at $2\theta = 24.90°$, approaching the $\gamma'$ reference peak position. Concurrently, the $(002)$ peak (panel (a3)) remains largely unchanged.

Annealing at $873$ K results in a further leftward shift of the $(001)$ peak to $2\theta = 24.87°$, and a pronounced sharpening ($\mathrm{FWHM} = 0.33°$) compared to the diffraction pattern at $673$ K, indicating reduced lattice distortion. The $(002)$ peak becomes asymmetric and can be deconvoluted into two distinct peaks centered around $50.97°$ and $51.57°$. The peak at $2\theta = 50.97°$ is close to the $\gamma'$-$(002)$ reflection at $51.18°$, while the peak at $2\theta = 51.57°$ is close to that of the as-deposited $Ni_{0.89}Al_{0.11}$ sample (Fig.1), i.e., it corresponds to the FCC $\gamma$-$Ni_{1-x}Al_x$ solid-solution phase.

Annealing at $1073$ K causes the $(001)$ peak to broaden again ($\mathrm{FWHM} = 0.74°$) and shift slightly rightward to $24.92°$, approaching the bulk $\gamma'$-$Ni_3Al$ $(001)$ peak at $24.94°$. While the $(002)$ peak remains asymmetric, the $\gamma'$ component on the left-hand-side shoulder becomes too broad to identify accurately, and the $\gamma$ component shifts to $2\theta = 51.65°$, closer to the angular position of the $(002)$ reflection in the as-deposited $Ni_{0.93}Al_{0.07}$ sample ($2\theta = 51.70°$) in Fig.1. This observation suggests that the decomposition of the nano-solution structure into distinct $\gamma$ and $\gamma'$ phases increases with annealing temperature. Furthermore, a weak peak is observed in the (111) reflection region, positioned between that of bulk $\gamma'$-$Ni_3Al$ $(111)$ $43.91°$ and Ni (111) $44.50°$.

Figures 5(b1) and (b2) present reduced module $\mathrm{Y}$ and hardness $\mathrm{H}$ values, respectively, of as deposited ($473$ K) films with composition $Ni_{0.93}Al_{0.07}$ (blue triangles), $Ni_{0.86}Al_{0.14}$ (black squares), and $Ni_{0.76}Al_{0.24}$ (red circles), as well as values for the sample with composition $Ni_{0.86}Al_{0.14}$ after annealing at $673$, $873$, and $1073$ K (black squares). The values are obtained from NI measurements at a maximum penetration depth of $40$ nm, with qualitatively similar trends found for NI data collected at $60$, $80$ and $100$ nm penetration depth (Section S5 and Fig. S3 in the Supporting Information [41]).

In Fig. 5(b1) we see that for all Al atomic fractions x ($0.07$, $0.14$, $0.24$) Y values in the range $220 - 240$ GPa are measured with overlapping error bars. Annealing of the $Ni_{0.86}Al_{0.14}$ sample to $673$ and $873$ K does not change Y significantly ($244 \pm 28$ and $223 \pm 19$ GPa, respectively), while a decrease of Y to $179 \pm 17$ GPa is observed after annealing at $1073$ K. With respect to the film hardness (Fig. 5(b2)), we find that increase of Al atomic fraction x from $0.07$ to $0.14$ causes H to increase from $3.1 \pm 0.3$ to $5.5 \pm 0.3$ GPa, while it remains almost constat ($5.4 \pm 0.4$ GPa) with further increase of x to $0.24$. Moreover, annealing of the $Ni_{0.86}Al_{0.14}$ sample at $673$ K leads to an increase of hardness $6.0 \pm 0.3$ GPa, followed by a decrease to $4.8 \pm 0.2$ and $4 \pm 0.3$ GPa after annealing at $873$ and $1073$ K, respectively.

Figure 5(c) shows top-view SEM images, EBSD and EDS data from the as-deposited and annealed ($873$ and $1073$ K) $Ni_{0.86}Al_{0.14}$ sample. The elemental distribution remains homogeneous across all annealing temperatures, as shown in panels (c3), (c4), (c7), (c8), (c10), and (c11). The EBSD analysis indicates a single out-of-plane crystallographic orientation for the as deposited sample (panel (c2)) and after annealing at $873$ K (panel ((c6)). Whereas annealing at $1073$ K (panel (c10)) shows the formation of a large $(111)$ grain which is consistent with the $(111)$ reflection in panel (a1). Notably, no elemental separation is observed within the $(111)$ grains.

By evaluating the data presented in Fig. 5 in relation to the structure formation discussed in Sections 3.1 and 3.2, we propose a pathway for explaining the evolution of microstructure and mechanical properties in magnetron sputtered $Ni_{1-x}Al_x$ films, as a function of composition and annealing temperature. In the as-deposited state, films with compositions $Ni_{0.76}Al_{0.24}$ and $Ni_{0.93}Al_{0.07}$ consist of $\gamma$ and $\gamma'$ phases, respectively, as suggested by XRD analysis and DFT/MS calculations. It is well known that the γ' phase serves as a strengthening agent in Ni-Al alloys [64] , which aligns well with our results, as the $Ni_{0.76}Al_{0.24}$ exhibits a greater hardness than the $Ni_{0.93}Al_{0.07}$ one ($5.4 \pm 0.4$ vs. $3.1 \pm 0.3$ GPa). Concurrently, the sample with composition $Ni_{0.86}Al_{0.14}$ exhibits a hardness comparable to that of the $Ni_{0.76}Al_{0.24}$ sample, despite its lower Al and thereby $\gamma'$-phase content. We suggest that this is due to the nano-solution structure, in which lattice distortion and strain increase the resistance of the material to plastic deformation.

Careful inspection of the XRD data in Figs. 5 (a1)-(a3) shows that the $(002)$ peak (panel (a3)) remains unchanged with respect to its shape and position when the $Ni_{0.86}Al_{0.14}$ sample is annealed at $673$ K, whereas the $(001)$ (panel (a1)) peak shifts toward the $Ni_3Al$ position, i.e., the spacing of the $(001)$ lattice planes within the $\gamma'$ domains approach its equilibrium value. This is a sign that $\gamma'$-phase domains grow in size and their lattice parameter relaxes, in agreement with previous studies on thermal treatment of Ni-Al alloys [65-67], Such microstructural evolution is consistent with the formation of distinct $\gamma$ and $\gamma'$ domains with different lattice parameters separated by strained domain boundaries. The lattice mismatch at the domain boundaries can hinder dislocation motion and explain the increase of hardness to the value of $6.0 \pm 0.3$ GPa (also referred to as age hardening). The split of the $(002)$ peak when the $Ni_{0.86}Al_{0.14}$ sample is annealed at $873$ and $1073$ K (Fig.5(a3)) shows that the nano-solution structure is decomposed and the $\gamma'$ and $\gamma$ phase domains coarsen further. This is consistent with a reduction of the domain boundary number density and the overall state in the film which can explain the decrease of the hardness to $4.8 \pm 0.2$ GPa ($873$ K) and $4 \pm 0.3$ GPa ($1073$ K).

The overall conclusion of the data presented in Section 3 is that the far-from-equilibrium nature of the sputtering process results in the formation of a nano-solution structure in the $Ni_{1-x}A_x$ films for x values within the range $0.11$ to $0.24$. In this structure, $\gamma$ and $\gamma'$ phase domains are intermixed at the atomic- and nanoscale forming a coherent lattice that enhances the alloy hardness relative to alloys with composition $x = 0.24$. The nano-solution is metastable as it dissolved into discrete $\gamma$ and $\gamma'$ domains when annealed at temperature $\geq 873$ K, while an age-hardening behavior is observed for annealing at $673$ K.

**4. Summary and conclusions**

$Ni_{1-x}Al_x$ thin alloy films with Al atomic fraction x from $0$ to $0.24$ are synthesized by magnetron co-sputtering on MgO$(001)$ at $473$ K. XRD data augmented by TEM show that at all compositions the films grow epitaxially despite the large lattice mismatch with the substate. Detailed experimental and theoretical analyses of the film microstructure show the formation of an FCC $Ni_{1-x}Al_x$ phase ($\gamma$ phase) for $x < 0.11$. Whilst for $0.11 \leq x \leq 0.24$ the microstructure comprises $\gamma$-phase domains and domains of the $L1_2$ $Ni_3Al$ $\gamma'$ phase. The two types of phase domains are intermixed at the atomic scale and are self-organized in a $\gamma/\gamma'$ nano-solution structure. The structure forms a coherent strained lattice which has a single lattice parameter and enhanced hardness in relation to single-phase structures. The overall microstructure is metastable, as evidenced by the shift and splitting of XRD peaks and the non-monotonic changes in the reduced modulus and hardness values upon annealing at temperatures in the range of $673-1073$ K. This study shows that the measurable lattice parameter of γ' phase can be systematically altered by the surrounding γ phase, thereby introducing a novel method for the structural design of the γ'/γ metastable intermetallic alloys. Further experiments could be carried out to investigate whether similar structures can be found in other intermetallic systems, as well as their potential applications. Moreover, the methodology presented herein can provide model material systems for shedding light onto fundamental processes pertaining to structure formation upon deposition and annealing (e.g., spinodal decomposition vs. nucleation and growth) —by combining of atomistic simulations and nanoscale probes—in intermetallic alloys that exhibit miscibility gaps in their equilibrium phase diagrams.

**Acknowledgements**

ZC acknowledges financial support by the Vilho, Yrjö and Kalle Väisälä Fund. KS acknowledges financial support by the Swedish Research Council (Grant No. VR-2021-04113) and the Åforsk foundation (Grant No. 22-150). DG acknowledges financial support from the Swedish Research Council (VR) through Grant No. 2023-00208. FT acknowledges financial support from the Research Council of Finland (grant nr 349602). EL acknowledges financial support from the Research Council of Finland (grant nr 354777). We acknowledge the OtaNano Nanomicroscopy Center for providing access to TEM, FIB, and XRD instrumentation. We thank Dr. Xuan Meng from Lanzhou University for his valuable contribution to TEM analysis.

## Figures and figure captions

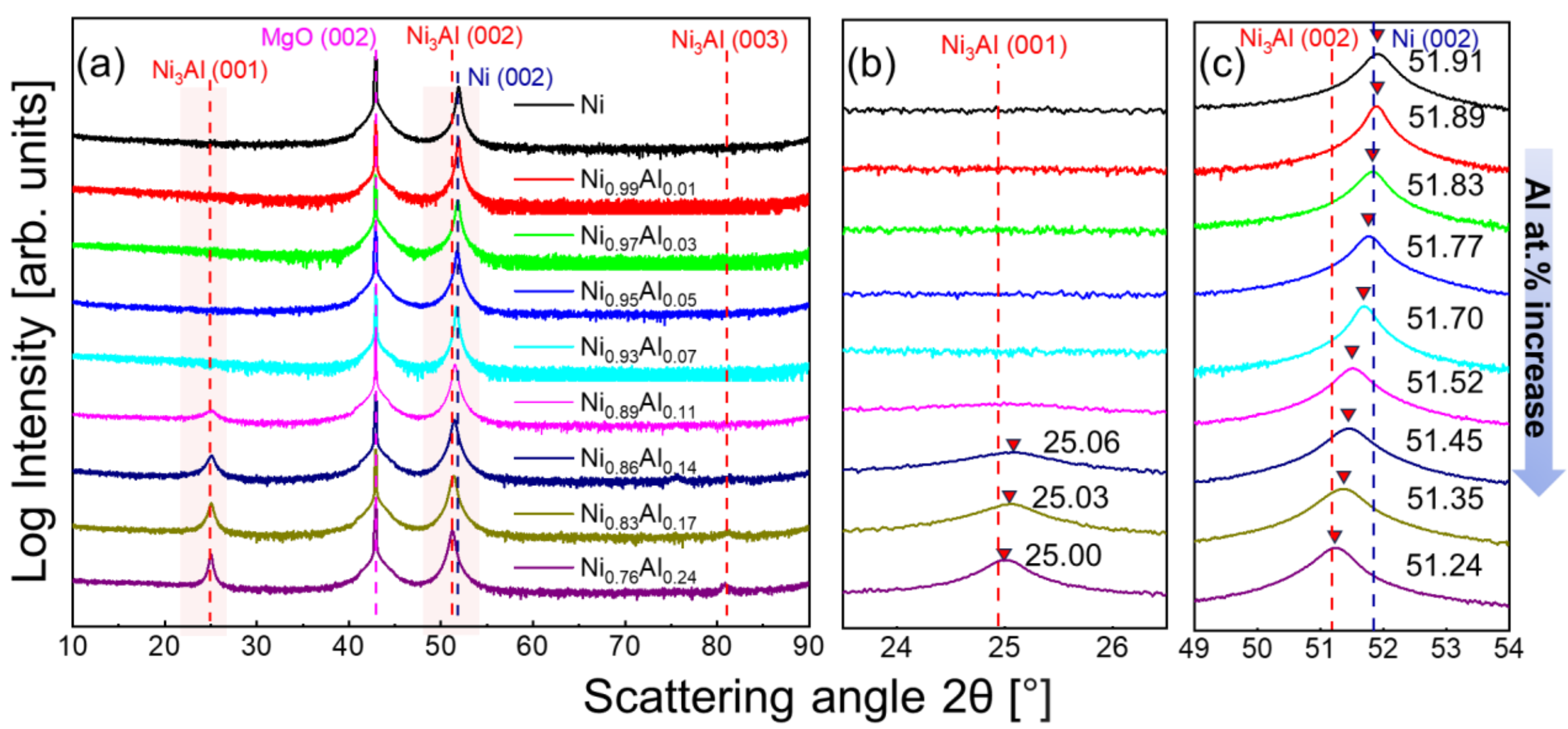

Fig.1. XRD patterns of Ni and $Ni_{1-x}Al_x$ films. Panel (a) shows patterns in the $2\theta$ angle range $10-90°$, while panels (b) and (c) present magnified regions in the $23.5-26.5°$ and $49-54°$ ranges, respectively. The vertical dashed lines in all panels indicate the angular position of reflections for the bulk unstrained MgO, Ni, and $Ni_3Al$. In (b) and (c) red arrows mark the peaks in the respective diffraction patterns, with the angular position provided next to each pattern.

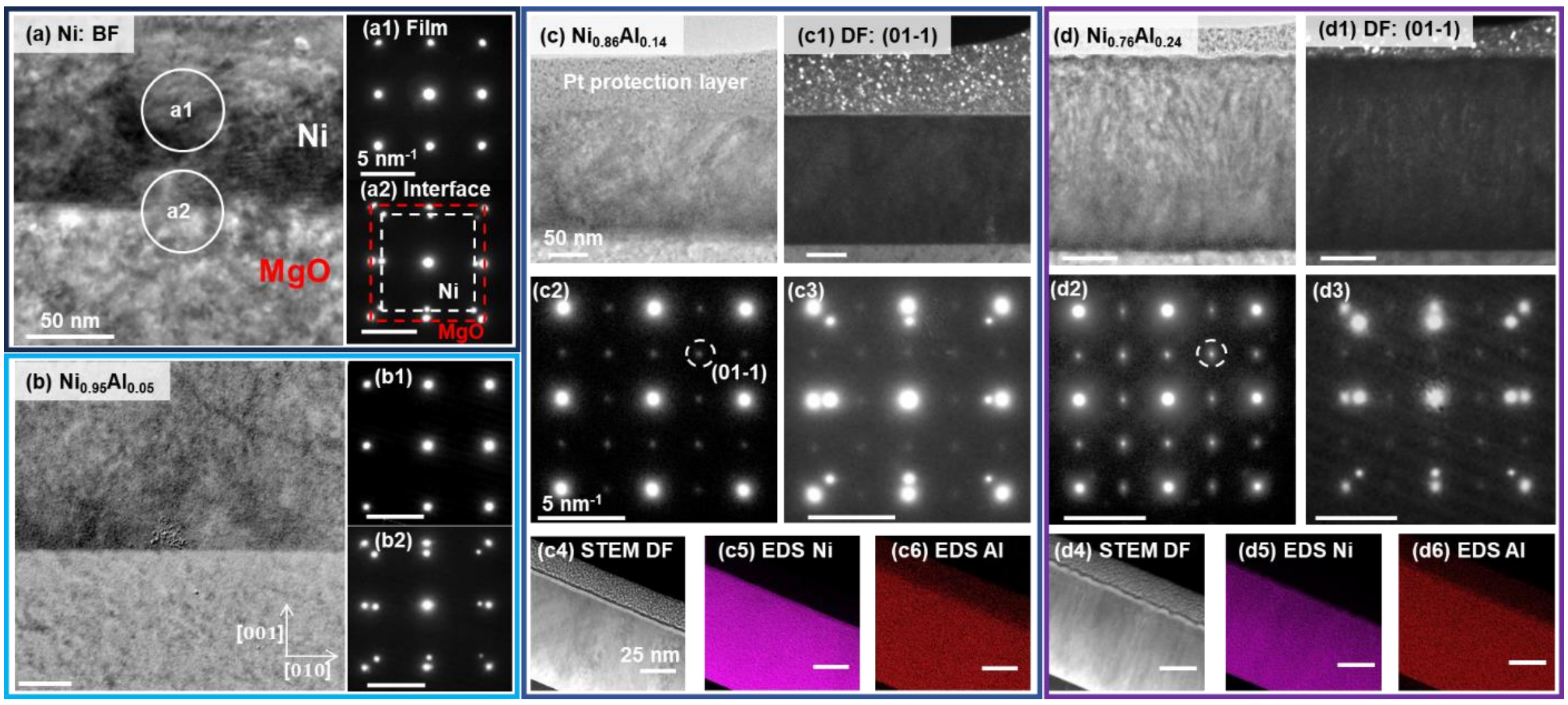

Fig.2. TEM images, SAD patterns and elemental distribution of $Ni_{1-x}Al_x$ thin films for $x = 0, 0.05, 0.14, \text{and } 0.24$. (a), (b), (c and c1), (d and d1) BF and complementary DF images of the films and interfaces (the scale bar is 50 nm in all TEM images). (a1 and a2), (b1 and b2), (c2 and c3), and (d2 and d3) SAD patterns of the corresponding film and interface regions along the $[100]$ zone axis, with a scale bar of $5\ \mathrm{nm}^{-1}$. The selected areas are marked with white circles in (a), and all other samples have similar positions. The SAD patterns from the film (Ni) and the substrate (MgO) are marked by white and red dashed squares in (a2). (c1), (d1) DF micrographs from $Ni_{0.86}Al_{0.14}$ and $Ni_{0.76}Al_{0.24}$ films, respectively, recorded using the $(01\bar{1})$ superlattice reflection of the $Ni_3Al$ structure (marked with a dashed white circle in (c1) and (d1). (c4-c6) and (d4-d6) show annular dark-field (ADF) scanning TEM (STEM) images and EDS elemental maps of Ni and Al in the $Ni_{0.86}Al_{0.14}$ and $Ni_{0.76}Al_{0.24}$ samples respectively.

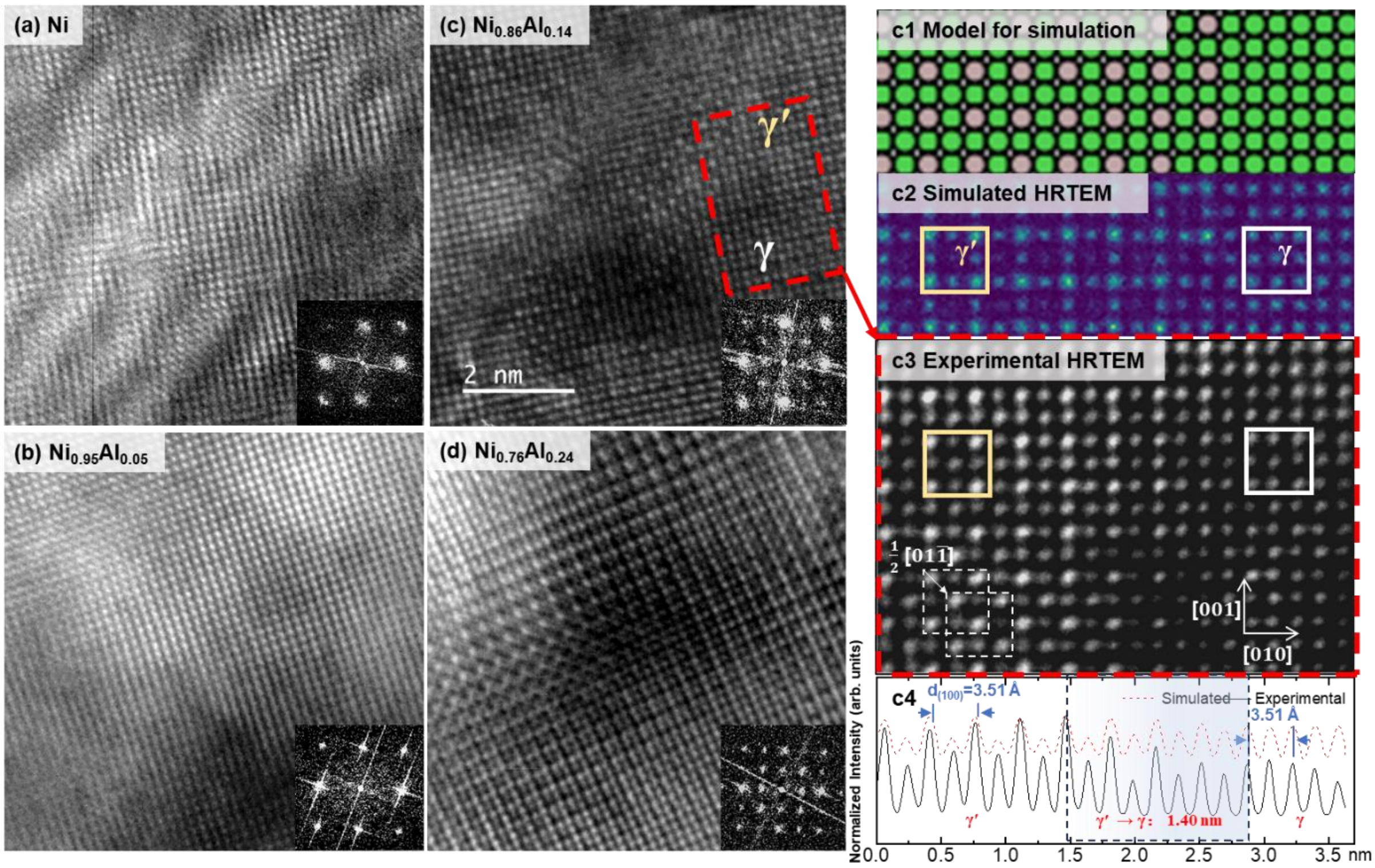

Fig.3. HRTEM images and FFT analysis of $Ni_{1-x}Al_x$ films compared to simulated images along the $[100]$ zone axis. Panels (a-d) display HRTEM images of $Ni_{1-x}Al_x$ thin films for $x = 0, 0.05, 0.14, \text{and } 0.24$, respectively, along with corresponding full-image FFT insets. The scale bar for all images is 2 nm. Panel (c3) shows an enlarged HRTEM image from panel (c), compared with a simulated HRTEM image (c2) and the corresponding model of a $Ni_3Al$ and Ni mixed super-cell (c1). The yellow solid, white solid, and dashed white boxes indicate the $\gamma'$, $\gamma$, and $\gamma'$ phases shifted along the $\frac{1}{2}$ $[01\bar{1}]$ direction, respectively. Panel (c4) compares the integrated intensity profiles along the $\gamma'$ and $\gamma$ phase domains from the experimental (c1) and simulated (c2) images. The scale of panels (c1-c4) is indicated on the horizontal axis in panel (c4).

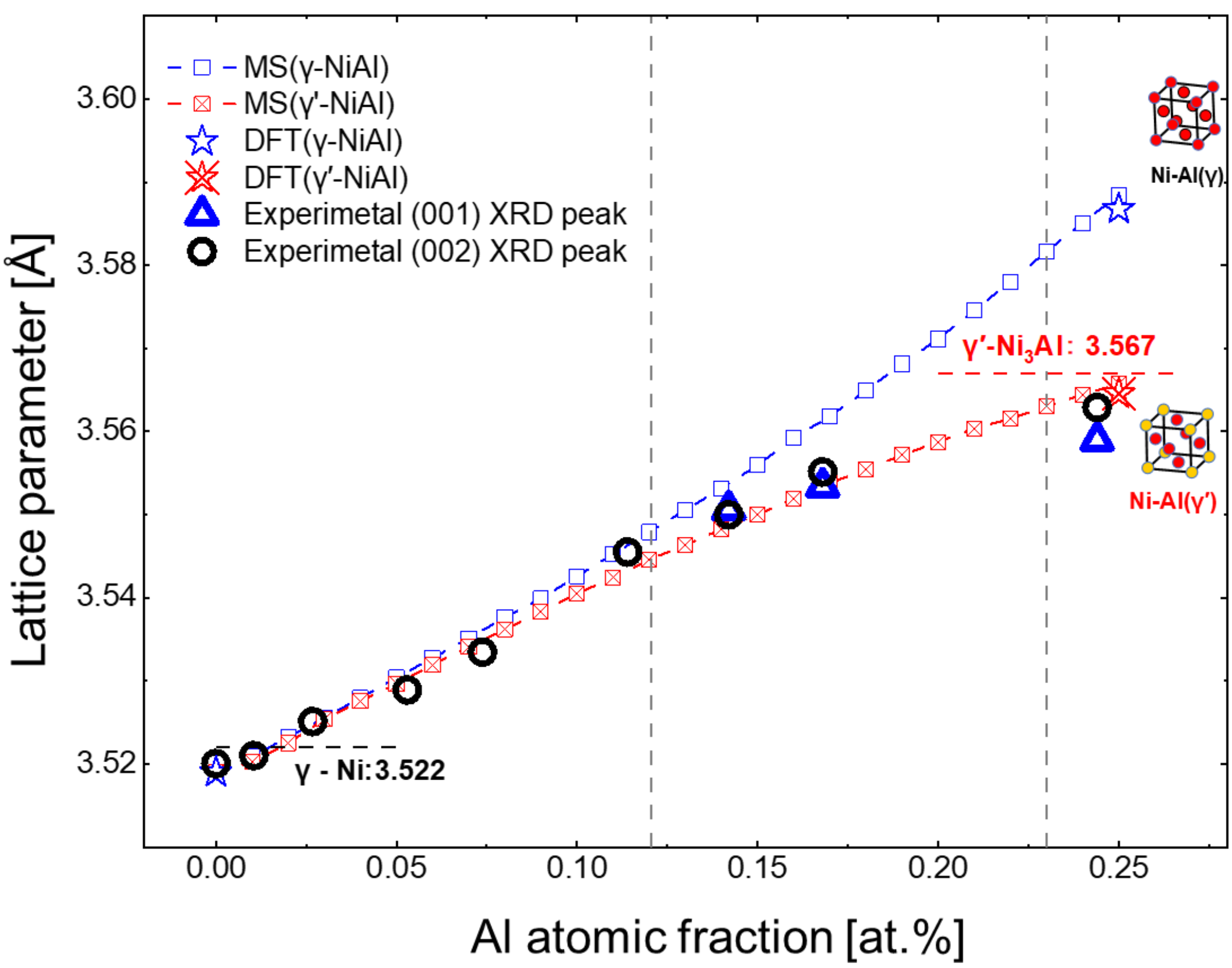

Fig.4. Lattice parameter of the Ni1-xAlx films vs. Al content x, as measured experimentally and determined from DFT and MS calculations. The blue triangles and black circles correspond to lattice parameter values determined from the angular positions of (001) and (002) XRD peaks (Fig.1), respectively. The blue and red squares show the lattice parameters determined from MS calculations for alloys with the γ random solid-solution and the ordered γ′ structure, respectively. The blue and red stars correspond to lattice parameter values determined by DFT for alloys with the γ and γ′ structure, respectively. The dashed lines connecting the data points from MS calculations are guides to the eye only.

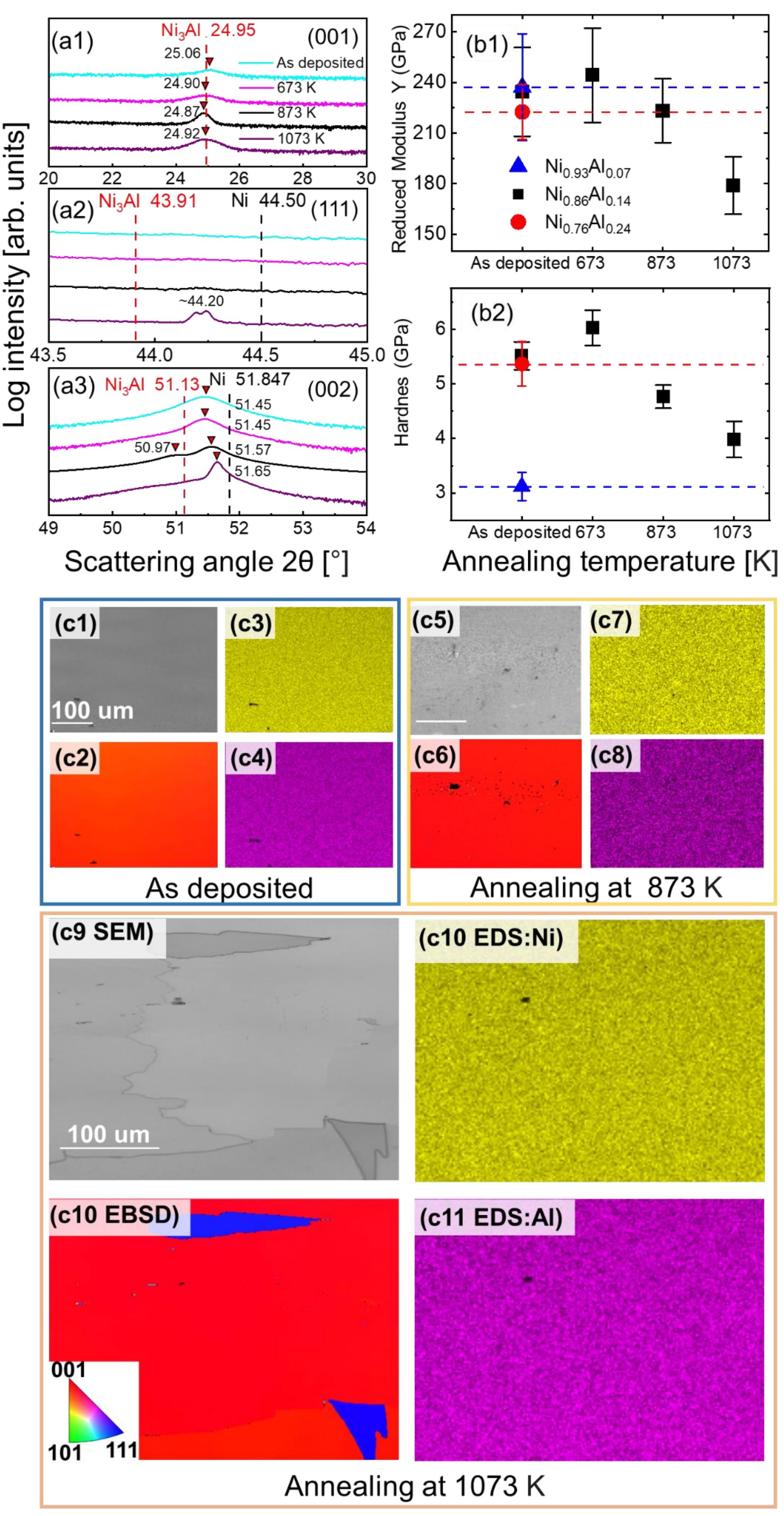

Fig.5. (a1-a3) XRD patterns of as-deposited and annealed at $673$, $873$ and $1073$ K $Ni_{0.86}Al_{0.14}$ films. Panels (a1), (a2) and (a3) present magnified regions around the $(001)$, $(111)$ and $(002)$ reflections for $Ni_3Al$ and Ni. The vertical dashed lines in all panels indicate the angular position of reflections in bulk and unstrained Ni and $Ni_3Al$. Red arrows mark the peaks in the respective diffraction peaks, with their angular position provided next to each pattern. Panels (b1) and (b2) show the reduced modulus $Y$ and hardness $H$ measured by nanoindentation (penetration depth $40$ nm) of annealed $Ni_{0.86}Al_{0.14}$ (black squares), as well as for as deposited $Ni_{0.93}Al_{0.07}$ and $Ni_{0.76}Al_{0.24}$ samples (blue and red-squares, respectively). Panels (c1) through (c12) present results from surface, elemental, and microstructure characterization before and after annealing of the $Ni_{0.86}Al_{0.14}$ sample. (c1), (c5) and (c9) Top-view scanning electron microscopy (SEM) images. (c2), (c6) and (c10) Electron backscatter diffraction (EBSD) inverse pole figure map. (c3), (c7) and (c11) and (c4), (c8) and (c12) Energy-dispersive X-ray spectroscopy (EDS) maps showing elemental distribution of Nickel (Ni) and Aluminum (Al).